\pdfoutput=1
\RequirePackage{rotating}
\documentclass[fleqn,usenatbib]{mnras}
\usepackage{newtxtext,newtxmath}
\usepackage[T1]{fontenc}
\DeclareRobustCommand{\VAN}[3]{#2}
\let\VANthebibliography\thebibliography
\def\thebibliography{\DeclareRobustCommand{\VAN}[3]{##3}\VANthebibliography}

\usepackage{graphicx}
\usepackage{amsmath}
\usepackage{soul}

\usepackage{url}
\usepackage{xcolor}
\usepackage[ruled,linesnumbered]{algorithm2e}
\usepackage{booktabs}

\usepackage{pdflscape}

\title[Catalog of Young Stellar Objects]{Deep Near-Infrared Survey Toward the W40 and Serpens South Region in Aquila Rift: 
A Comprehensive Catalog of Young Stellar Objects}

\author[J. Sun et al.]{
Jia Sun,$^{1,2}$\thanks{E-mail: jiasun@pmo.ac.cn or sun.jiaaaa@gmail.com}
Robert A. Gutermuth,$^{3}$
Hongchi Wang,$^{1}$
Miaomiao Zhang,$^{1}$
Shuinai Zhang,$^{1,4}$ 
\newauthor
Yuehui Ma,$^{1}$
Xinyu Du,$^{1}$
Min Long$^{5}$
\\
$^{1}$Purple Mountain Observatory, Chinese Academy of Sciences,
No.10 Yuanhua Rd, Qixia District, Nanjing 210033, China\\
$^{2}$University of Chinese Academy of Sciences, No.19(A) Yuquan Rd, Shijingshan District, Beijing 100049, China\\
$^{3}$Department of Astronomy, University of Massachusetts, Amherst, MA 01003, USA\\
$^{4}$Key Laboratory of Dark Matter and Space Astronomy, Chinese Academy of Sciences, China\\
$^{5}$Department of Computer Science, Boise State University, USA
}

\date{Accepted for publication in MNRAS}

\pubyear{2022}

\begin{document}
\label{firstpage}
\pagerange{\pageref{firstpage}--\pageref{lastpage}}
\maketitle

\begin{abstract}
Active star forming regions are excellent laboratories for studying the origins and evolution of young stellar object (YSO) clustering.
The W40 - Serpens South region is such a region, and we compile a super near-and-mid-infrared catalog of point sources in it, based on deep NIR observations of CFHT in combination with 2MASS, UKIDSS, and Spitzer catalogs.
From this catalog, we identify 832 YSOs, and classify 15, 135, 647, and 35 of them to be the deeply embedded sources, Class I, Class II YSOs, and transition disk sources, respectively.
In general, these YSOs are well correlated with the filamentary structures of molecular clouds, especially the deeply embedded sources and the Class I YSOs.
The W40 central region is dominated by Class II YSOs, but in the Serpens South region, a half of the YSOs are Class I.
We further generate a minimum spanning tree (MST) for all the YSOs.
Around the W40 cluster, there are eight prominent MST branches that may trace vestigial molecular gas filaments that once fed gas to the central natal gas clump.
Of the eight, only two now include detectable filamentary gas in Herschel data and corresponding Class I YSOs, while the other six are exclusively populated with Class II.
Four MST branches overlap with the Serpens South main filament, and where they intersect, molecular gas ``hubs'' and more Class I YSOs are found.
Our results imply a mixture of YSO distributions composed of both primordial and somewhat evolved YSOs in this star forming region.
\end{abstract}


\section{Introduction} \label{chap:Intro}
Stars are the fundamental building blocks of galaxies.
They originate from collapse of molecular clouds, the coldest and densest part of the interstellar medium (ISM), and mainly form in over-dense groupings and clusters within those clouds.
The forming stars can provide significant feedback to surroundings and ultimately affect the formation and evolution of subsequent generations of stars. 
The deaths of high mass stars in supernovae can have profound effects on nearby ISM conditions. 
Thus, stellar evolution from birth to death influences the evolution of multiple objects: stellar systems, molecular clouds, and host galaxies. 
The understanding of current star formation is critical to our growing knowledge of galaxy formation and evolution, especially those regions containing young stellar objects (YSOs) at differential evolving stages. 

YSOs usually reside in dusty molecular cloud cores, which make them dim in the optical band or even undetectable.
Nevertheless, they are visible at near-, mid-, and far-infrared (NIR, MIR, and FIR) bands, not only because of their photosphere blackbody spectrum but also the overwhelmed dust emission in their circumstellar materials.
Therefore, at different evolving stages, YSOs have distinct infrared spectral energy signatures \citep{lada1987, andre2002}.
Meanwhile, they also emit occasionally in the radio band due to free-free emission from an ionized plasma \citep[e.g.,][]{shang2004}, in the UV band due to accretion shock or a strong stellar wind \citep[e.g.,][]{calvet2004}, and in the X-ray band due to flares \citep{preibisch1999,tsuboi2001,winston2018}.

During the cloud evolution and star formation, there are a few observable stages of YSOs after the starless core stage: 
(1) Class 0 sources,  which are deeply embedded protostellar condensations that can be most effectively identified at sub-millimeter, millimeter, and FIR bands due to the dust blackbody emission \citep{hurt1996, testi1998, bontemps2010}. 
(2) The canonical category of Class I YSOs, including flat spectrum objects, describes protostars that have accreted enough mass and started to form circumstellar disks, but still possess infalling dusty envelopes \citep{caratti2017, greene1994, fang2009}.
(3) For a Class II YSO at the "disky" pre-main sequence star stage, the envelope is largely cleared and an optically thick protoplanetary disk remains behind \citep{liu2019}.
(4) Class III YSOs have started the hydrogen nuclear fusion, whose disks have only debris or some planets left \citep{su2006}. 
At last, with increasing temperature, the YSOs come into the main sequence.

Sensitive large-scale sky surveys at infrared wavelengths can provide a powerful tool to generate comprehensive and high confidence catalogs of YSOs, especially in nearby (within 1 kpc) star forming regions.  
Such surveys include “From Molecular Cores to Planet” (c2d) \citep{evans2009} and “Gould Belt Survey” (GB) \citep{dunham2013} conducted by the Spitzer Space Telescope, and “Two Micron All Sky Survey” (2MASS) \citep{skrutskie2006}.  
Based on these surveys, thousands of YSOs have been identified and classified in many nearby star forming regions \citep[e.g.,][]{gutermuth2009,dunham2015}.

This work studies two conjuncted star-forming regions among those, namely, W40 and Serpens South (Fig. \ref{fig:control_target_demo}), both of which belong to a larger complex of molecular clouds collectively known as the {\it Aquila Rift}. 
The W40 region has been observed in the NIR band since 1978 \citep{zeilik1978}, but only after 30 years the Serpens South region was discovered and named \citep{gutermuth2008serpens}.
The W40 region is famous for a hourglass-shaped H II region illuminated by a few massive OB stars \citep{shuping2012}, and its cloud structure, magnetic field, and star-formation activities such as outflows and YSO populations have been studied in detail through observations from NIR to millimeter bands \citep{gutermuth2008serpens, nakamura2011, sugitani2011, zhang2015, winston2018, shimoikura2019}.
The Serpens South region is embedded in a dark cloud filament and contains a large number of extremely young objects, which received a lot of attention too.
The two regions are only separated by $\sim$20$'$ (or $\sim$3 pc) from each other, and both of them are $\sim$436 pc \citep{ortiz-leon2017} away.
As a consequence, the W40 - Serpens South region can be considered as one complex with YSOs at different states, presenting an excellent laboratory to study the feedback influences.

In this region, \cite{bontemps2010} has used the Herschel Gould Belt Survey data to identify several tens of Class 0/I sources at very early evolutionary stages. \cite{dunham2015} has used the c2d and Spitzer GB Survey data to identify YSOs, with NIR data based on the 2MASS survey which is not deep enough for those embedded in thick molecular filaments.
In this work, we analyze NIR data provided by the Canada-France-Hawaii Telescope  (CFHT) for the W40 - Serpens South region, complemented by catalogs such as 2MASS and UKIDSS.
The deeper NIR observations can penetrate better toward the filaments around W40 and Serpens South cores, and allow us to identify more embedded YSOs and depict detailed structures of young clusters. 
Although we do not include a specific filament analysis here, we want to briefly check the spatial association between YSOs and filaments in the overall W40 - Serpens South region.
Such YSO structures can also be compared with a more exquisite structure of molecular clouds for example be detected by ALMA \citep{plunkett2018}.

The paper is organized as follows.
The used observations and data are introduced in Section \ref{chap:Obs}. 
Methods of image processing and point source catalog production are explained in Section \ref{chap:dataprocess}. 
We present the identification of YSOs and discuss their relationships with molecular clouds and the entire star forming region in Section \ref{chap:id_YSOs} and include the summary in Section \ref{chap:sum}.

\section{Observations and Data}\label{chap:Obs}
\subsection{CFHT Observations} 

\begin{figure}
\centering
\includegraphics[width=\linewidth]{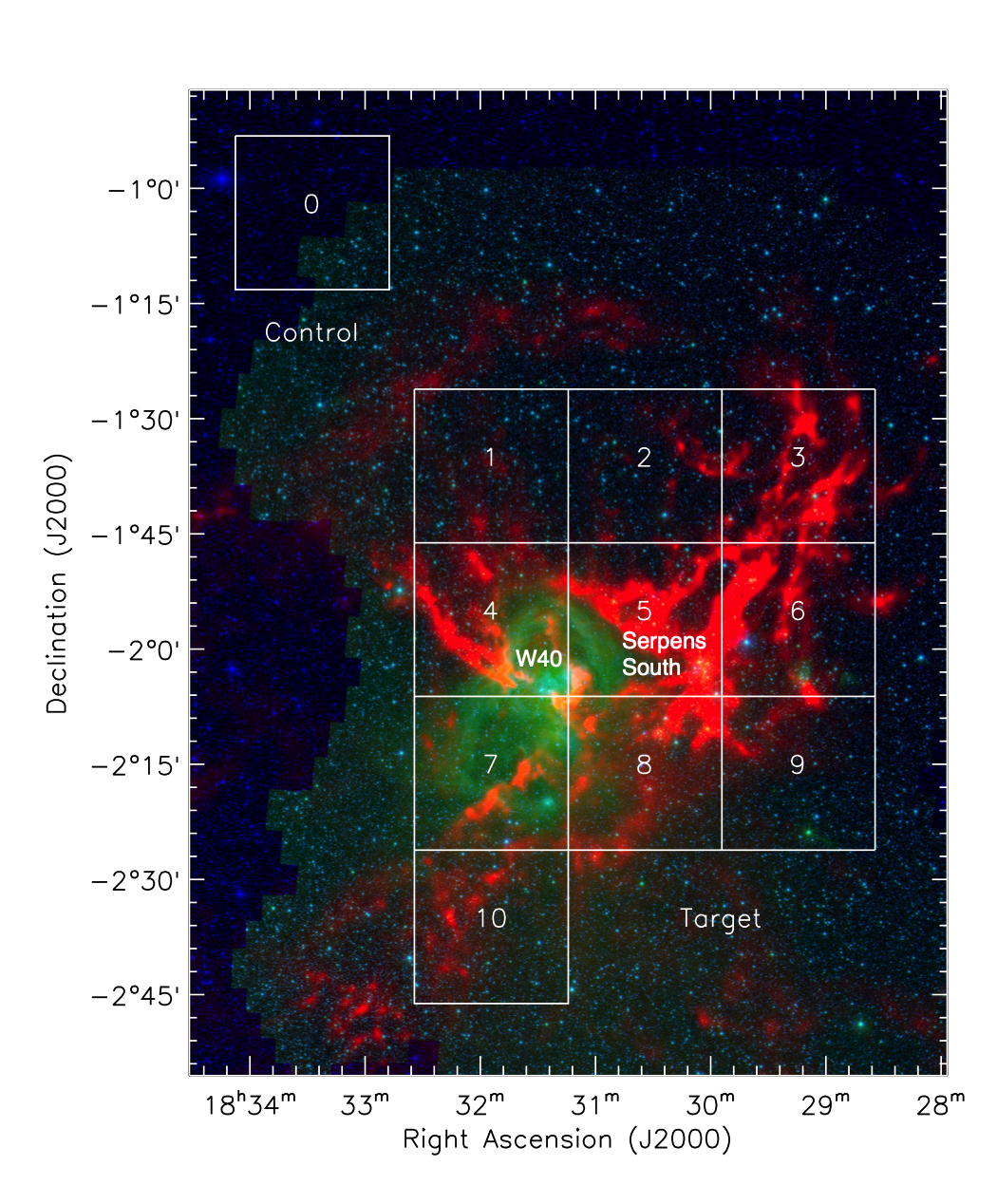}
\caption{Composite IR image of W40 and the Serpens South.
White box: ``Control'' (No. 0) and ``Target'' (No. 1-10) fields in WIRCam. 
Underlaid image is composed of 2MASS K$_{\rm s}$ (blue), Spitzer 4.5 $\mu$m (green) and Herschel hydrogen column density (red). 
\label{fig:control_target_demo}}
\end{figure}

We use Wide-Field Infrared Camera (WIRCam) at the CFHT to take NIR images of the W40 and the Serpens South at the J, H, and K$_{\rm s}$ bands.
WIRCam has a field of view of $20' \times 20'$, and a resolution of 0.3$''$ per pixel.
The ``Target'' field has a size of $\sim 60' \times 80'$ or $\sim$ 7.6 pc $\times$ 10.1 pc and covers most of the main molecular complex.
Fig. \ref{fig:control_target_demo} shows the entire region consisting of 10 sub-fields of the ``Target'' labeled from No. 1 to 10 during the actual observation and one ``Control'' field (labeled No. 0) of $20' \times 20'$ as reference. 
The ``Control'' field is selected based on the distance to the``Target'', and has a very low density of molecular clouds, so there's no young stars enclosed. 
2MASS (blue), Spitzer 4.5 $\mu$m (green) and Herschel hydrogen column density (red) are also superposed for object identification. It can be seen that the 2MASS data covers the entire area and Spitzer data covers only the ``Target'' and $\sim$1/4 ``Control'' field, Herschel data covers only the ``Target'' field.

The CFHT observations were carried out in the J, H, and K$_{\rm s}$ bands between 2012 July 27 and 29. 
Seeing is good and the average full width at half-maximum (FWHM) of sub-fields varies from 0.$''$45 to 0.$''$96.
Single image exposure time of J, H, and K$_{\rm s}$ is 30, 10.3 and 15 s, respectively. 
We selected the dithering mode so that bad pixels and cosmic rays can be effectively removed. 
Each sub-field is covered by 4--24 exposures, and about 70\% of the final integrated images contain more than 16 exposures. 
Total exposure time reaches 240, 123.6, and 120 s on average, in the J, H, and K$_{\rm s}$ bands, respectively. 

The image of each individual exposure is processed by CFHT's IDL Interpreter of the WIRCam Images,  i.e., \texttt{`I`wii} pipeline \footnote{\href{http://www.cfht.hawaii.edu/Instruments/Imaging/WIRCam/IiwiVersion1Doc.html}{www.cfht.hawaii.edu/Instruments/Imaging/WIRCam/IiwiVersion1Doc.html}}. 
In the pipeline, following processes are performed automatically: chipbias and dark subtraction, flat-fielding, non-linearity correction, cross-talk removal and sky subtraction. 
Pixels that are bad or saturated are also marked.

\subsection{2MASS and UKIDSS Data}
Bright stars in our long exposure time images could be saturated and need to be resolved in our CFHT observations. 
This is also because of the wide field and high sensitivity the WIRCam has. 
For this reason, we combined both 2MASS and UKIDSS data for calibration of those bright stars. 

First, we use the 2MASS Point Source Catalog (PSC) \citep{skrutskie2006}. 
2MASS data corresponding to the ``Target'' and the ``Control'' fields are available from VizieR II/246 catalog\footnote{\href{http://vizier.u-strasbg.fr/viz-bin/VizieR?-source=II/246&-to=3}{http://vizier.u-strasbg.fr/viz-bin/VizieR?-source=II/246\&-to=3}}. 
We use stars with a high quality flag (\texttt{qflag} $=$ `AAA'), which means the source has a signal-to-noise (S/N) ratio greater than 10 and error less than 0.1. 
We also use a low contamination flag (\texttt{cflag} $=$ `000'), which means the source is unaffected by any known artifacts, or simply has no detection in some bands.

Second, we introduce UKIRT Infrared Deep Sky Survey (UKIDSS) Galactic Plane Survey (GPS) data \citep{lucas2008} for calibration and use it as the medium-bright part of our catalog. 
UKIDSS has integration time of 80 s, 80 s, and 40 s for J, H, K bands respectively.
The UKIDSS data we use is available at WFCAM Science UKIDSS Archive release DR11 \footnote{\href{http://wsa.roe.ac.uk:8080/wsa/SQL_form.jsp}{http://wsa.roe.ac.uk:8080/wsa/SQL\_form.jsp}}. 
Quality flags are selected as follows: \texttt{priOrSec} $=$ 0 or \texttt{frameSetID} for selecting out a seamless, best catalog; \texttt{ppErrbits} $<$ 256 to exclude all K band sources with any error quality condition; $-2 \le$ \texttt{Class} $\le -1$ to exclude noise, saturated and galaxy.
The last flag is strict, it cuts off $\sim$32\% of total UKIDSS stars.
Among them, sources marked to be ``galaxy'' are mostly pairs of stars \citep{lucas2008}, therefore we will use CFHT data instead.

\subsection{Spitzer Data}
MIR data is a good supplement to NIR for both contamination removal and YSOs identification. 
Aquila data from the Spitzer Gould Belt Legacy Survey are therefore included in this work. 
This section is summing up work performed by Gutermuth et al. 2022 in preparation.

The ``Target'' field is fully covered, and the ``Control" field is 1/4 covered by the Spitzer survey. 
The Spitzer photometry is derived from a draft catalog of the Spitzer Extended Solar Neighborhood Archive (SESNA;  R. Gutermuth et al. 2022, in preparation), and we summarize below their creation process via the Cluster Grinder IDL software collection for Spitzer data treatment \citep{gutermuth2009, winston2018, li2019, pokhrel2020}.

The \texttt{IRAC} imaging has four bands at 3.6, 4.5, 5.8, 8.0 $\mu$m and the \texttt{MIPS} imaging is at 24 $\mu$m \citep{gutermuth2008serpens}.  
Images are treated for bright source artifacts using custom IDL scripts. 
Cosmic ray hits are identified and masked out and all images are merged into mosaics using the WCS mosaic module, with 0.87$''$ per pixel resolution for \texttt{IRAC} and 1.8$''$ per pixel for the \texttt{MIPS}.

Sources are identified using the \texttt{PhotVis} module\citep{gutermuth2008ngc1333}, and aperture photometry of those sources is performed using \texttt{aper.pro} from the IDL Astronomy User's Library \citep{landsman1993}.
The adopted aperture radius for \texttt{IRAC} is 2.4$''$, and the background is estimated from an annulus field with radii of 2.4$''$ and 7.2$''$.  
Corresponding \texttt{MIPS} aperture and inner and outer annulus radii are 6.35$''$, 7.62$''$, and 17.78$''$, respectively.  
Instrumental fluxes are converted to Vega-standard magnitudes using magnitude zero points (e.g. magnitude for 1 DN/s net flux in aperture) of 20.21, 19.45, 17.25, and 17.64 mag for \texttt{IRAC} channels 1 through 4 and 15.35 mag for \texttt{MIPS} 24 $\mu$m.  
Fields averaged 90\% differential completeness magnitude limits are 14.5, 14.2, 13.2, 12.0, and 8.4 mag at 3.6, 4.5, 5.8, 8.0 and 24 $\mu$m, respectively.

\subsection{Herschel Data}
We also include the publicly available Herschel hydrogen column density map from the Herschel Gould Belt Survey for higher resolution of dust extinction estimation and better demonstration. 
The map is derived from greybody fitting of the 160, 250, 350, and 500 $\mu$m FIR images, and resampled and smoothed to match the latter image's 36$''$ FWHM beam size \citep{konyves2015}. 

\section{Image Reduction, Aperture Photometry and Super Infrared Catalog}\label{chap:dataprocess}
After 352, 528, and 384 WIRCam small images in J, H and K$_{\rm s}$ band are output by \texttt{`I`wii} pipeline, many corrections are performed before the final catalog is generated.
This section describes the reduction and analysis of the NIR data only.

\subsection{Follow-up Reduction of Images}\label{chap:follow-up-reduction}
Original images have many void pixels that affect both faint and bright stars, and combined images can effectively remove them and also give larger S/N for star detection.
Meanwhile, the resampling process can correct the distortions of original images.
To co-add individual images together, we choose a suite of astronomical pipeline softwares ``AstrOmatic'' \footnote{\href{https://www.astromatic.net/software/}{https://www.astromatic.net/software/}} which includes functions: SExtractor (Source Extractor), SCAMP (Software for Calibrating AstroMetry and Photometry), and SWarp (Source Warp) \citep{bertin2010se,bertin2010sc,bertin2010sw}.
Those tools are developed at Institut d'Astrophysique de Paris (IAP) in France and designed to be run in a batch mode on large quantities of data.

First, SExtractor extracts point source catalog from images and outputs a rough catalog. 
Second, SCAMP compares it to the standard 2MASS catalog and calculates an astrometric solution. 
For our data, SCAMP corrects the catalog distortion with polynomial functions under the TPV World Coordinate System \footnote{\href{https://fits.gsfc.nasa.gov/registry/tpvwcs/tpv.html}{https://fits.gsfc.nasa.gov/registry/tpvwcs/tpv.html}} .
With these equation coefficients, SCAMP outputs the corresponding external header to each image.
Eventually, SWarp does image resampling and co-addition using polynomial distortion correction terms in external headers. 
We observe this suite of tools performs better than SIMPLE Imaging and Mosaicking Pipeline (SIMPLE) \footnote{\href{http://group.asiaa.sinica.edu.tw/whwang/idl/SIMPLE/WIRCAM/Doc/simple.pdf}{http://group.asiaa.sinica.edu.tw/whwang/idl/SIMPLE/WIRCAM/Doc/ simple.pdf}}. 
In the SIMPLE, astrometry is done with more resampling, during which extra spatially correlated noise is introduced \citep {zhang2019} and FWHM of sources increase $\sim$10\%.

Detrended images output by CFHT \texttt{`I'iwi} pipeline are input to the above pipeline. 
In SExtractor and SCAMP we use all default settings. 
We change three parameters instead of using the default in SWarp: combine with ``AVERAGE'', resampled pixel scale to be``$0.15''$'', and background ``NOT'' subtracted. 
For our images ``AVERAGE'' is better than the default ``MEDIAN'' though the latter can effectively remove cosmic rays.
Although stars FWHM increase on the final images, the sources are shaping well for photometry, as``AVERAGE'' can better preserve the shape of a source, especially for the peak.
A 0.15$''$ pixel scale makes better image quality where pixels are not integrally overlapped and defines better where a star is located. 
Background is not subtracted because it contains nebular emission.
After SWarp, the ``Target'' field is stitched to 10 pieces of images instead of one, which is required by the astigmatism correction in section \ref{chap:astigmatism}. 
Therefore, the final catalog is also stitched from 10 separated catalogs. 
After distortion correction, the typical astrometry difference with 2MASS for a star is 0.06$''$.
In the final derived maps, the mean FWHM varies from 0.65$''$$\sim$1.0$''$, as images are more stretched after combining.

\subsection{Aperture Photometry} 
On final stacked images, point spread functions (PSF) are not uniform and their sizes in each field are not the same. 
So we choose to do aperture photometry instead of PSF (PSF) photometry. 

We use PhotVis \citep{gutermuth2008ngc1333} to identify stars and perform aperture photometry.
The radius of aperture, inner and outer radii of the background annulus are chosen to be 0.75$''$, 1.5$''$ , and 3$''$, which are relatively small to avoid nearby starlight pollution when the typical nearest neighbor distance is $\sim$1.8$''$ for a star.
We identify sources with statistics better than 6 S/N ratio. 
The numbers of sources from the initial photometry are 422,442 for J, 543,212 for H, and 813,307 for K$_{\rm s}$. 
The number of detections in the K$_{\rm s}$ bands is nearly twice with respect to those detected in the J band, mainly because that longer wavelength emitted by blocked background Milky Way stars can better penetrate dark clouds. 
We use TOPCAT to perform the cross-match, between K$_{\rm s}$ and H bands first, and then between K$_{\rm s}$/H and J bands, because stars at the first two bands have lower FWHM values and larger number of detections.
In total there are 870,471 sources identified, and 369,420 sources appear in all the three bands.
Position tolerance of cross-match of two catalogs is 1$''$. 

There are two methods to estimate detection completeness. 
The first method is to successively add synthetic stars with known magnitude and count their detection ratio, as described in \cite{gutermuth2005}. 
The second method is to fit a linear relationship between magnitude histogram of log(N) and magnitude, and then assume that the decline at faint magnitudes is entirely due to limited sensitivity \citep{lucas2008}. 
The latter method is easier in our case as synthetic stars change in those corrections from the first photometry to the final magnitude catalog. 
The typical 90 \% completeness limits in uncrowded fields are J $=$ 19.8, H $=$ 18.6, and K $=$ 18.0, with uncertainties to be $\sim$0.2 mag. 
The non-uniform number of exposures in the different subfields also leads to different limiting magnitudes within the subfields and to a non homogeneous photometric catalogue. We have checked that the mixed magnitude for the three bands is $\sim$0.12 mag less than the theoretical magnitude, which is consistent with our derived uncertainties.

\subsection{Effects of Telescope Spike and Long Time Exposure}

\begin{figure}
\centering
\includegraphics[width=\linewidth]{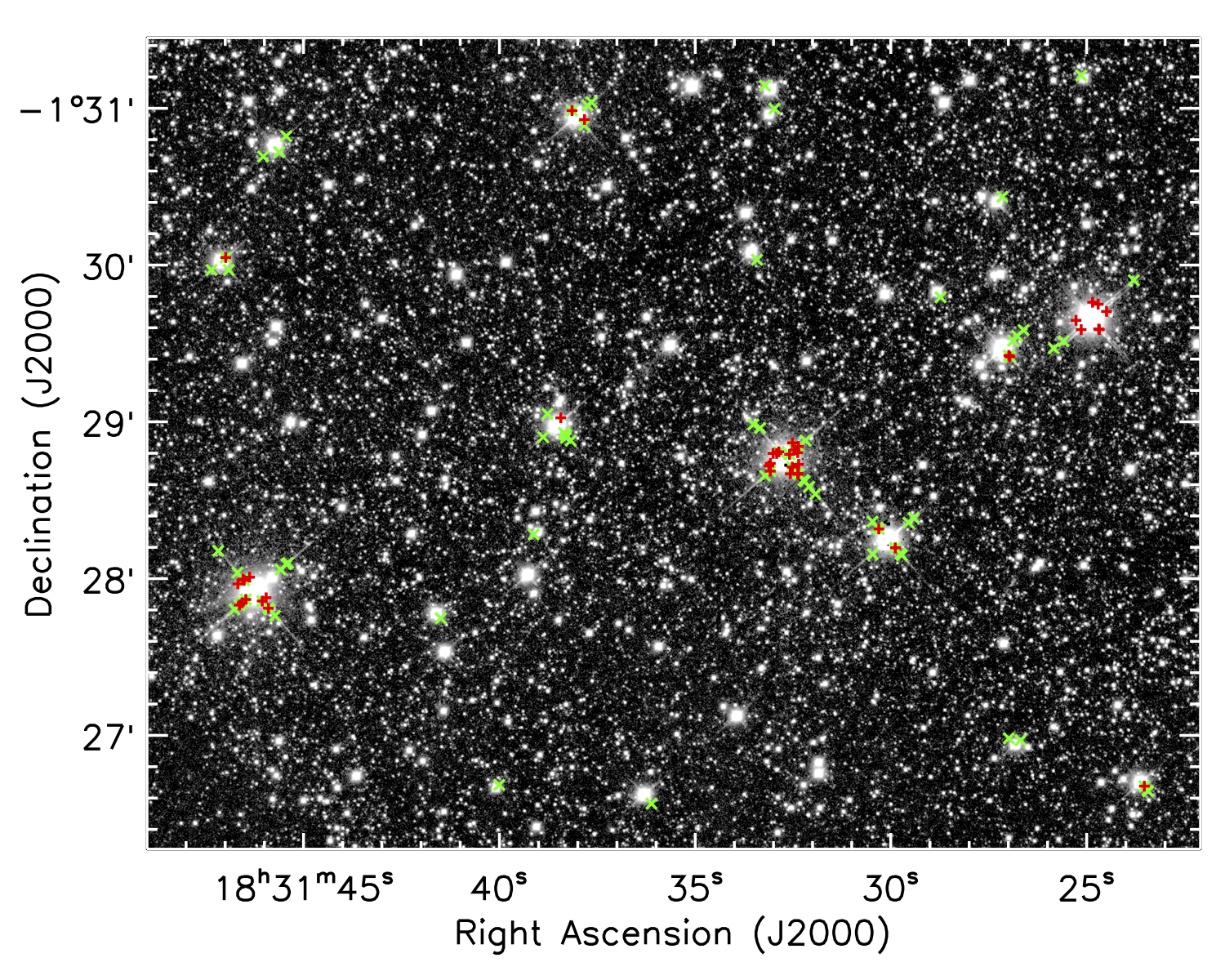}
\caption{
Effects of bright stars on faint stars due to telescope spikes and long exposure on J band image. 
Green cross signs are faint stars that get marked as ``affected by a spike''. 
Red plus signs are stars that are marked as ``affected by bright field''. 
\label{fig:spike_demo}}
\end{figure}

Bright stars may affect identification of faint stars due to effects of telescope spikes and long exposure.
Modern reflecting telescopes usually need spider vanes to support their second mirror. 
When incident light reaches diffraction limit, telescope spikes can appear near these bright stars. 
In our CFHT images, bright stars have four main spikes. 
In addition, after long exposure, the bright stars can look bigger and more square in shape and thus affect their surrounding field.
Fig. \ref{fig:spike_demo} shows both effects on J band images. 
We examine bright stars with J/H/K Mag $<$ 13.5 and mark affected faint stars. 

To quantify those effects and mark faint stars, we introduced two values: spike length $L_\mathrm{spike}$ and radius of affection ($R_\mathrm{affect}$).
First, We choose bright stars from the 2MASS catalog, and their magnitudes are evenly distributed. 
Second, we roughly record the pixel length of their spikes and their affected radius. 
By assuming an exponential relationship, we fit both $\log(L_\mathrm{spike})$ and $\log(R_\mathrm{affect})$ as a linear function of magnitude $M$. 
The resultant formulas are as follows:
\begin{align}
L_\mathrm{spike}(\mathrm{pixels})&=10^{a_{0}+a_{1}M},\\
R_\mathrm{affect}(\mathrm{pixels})&=10^{b_{0}+b_{1}M}.
\end{align}
The fitting parameters are shown in Table \ref{tab:LR}.
\begin{table}
\centering
 \caption{Fitting parameters for for spike length and radius of affection.}
 \label{tab:LR}
 \begin{tabular}{lcccc}
  \toprule
          Bands           & $a_0$   & $a_1$  & $b_0$   & $b_1$ \\
  \hline
         J                    & 3.91      & -0.206  & 3.30      & -0.182 \\
         H                   & 3.74      & -0.209  & 3.19      & -0.187 \\
K$_\mathrm{s}$     & 3.51      & -0.167  & 2.84      & -0.145 \\
 \bottomrule
 \end{tabular}
\end{table}
	
These functions are applied to our field to mark any star that is located nearby the bright sources.  
There are around 12,000 sources to be marked as ``affected by a spike'' and 4,700 sources to be marked as ``affected by a bright field''. 
Two of our final YSOs are marked as ``affected'', so we re-classify them with the 2MASS data instead. 
Coincidentally, their evolution stages remain unchanged.

\subsection{Astigmatism Correction}\label{chap:astigmatism}
\begin{figure*}
\centering
\includegraphics[width=0.95\linewidth]{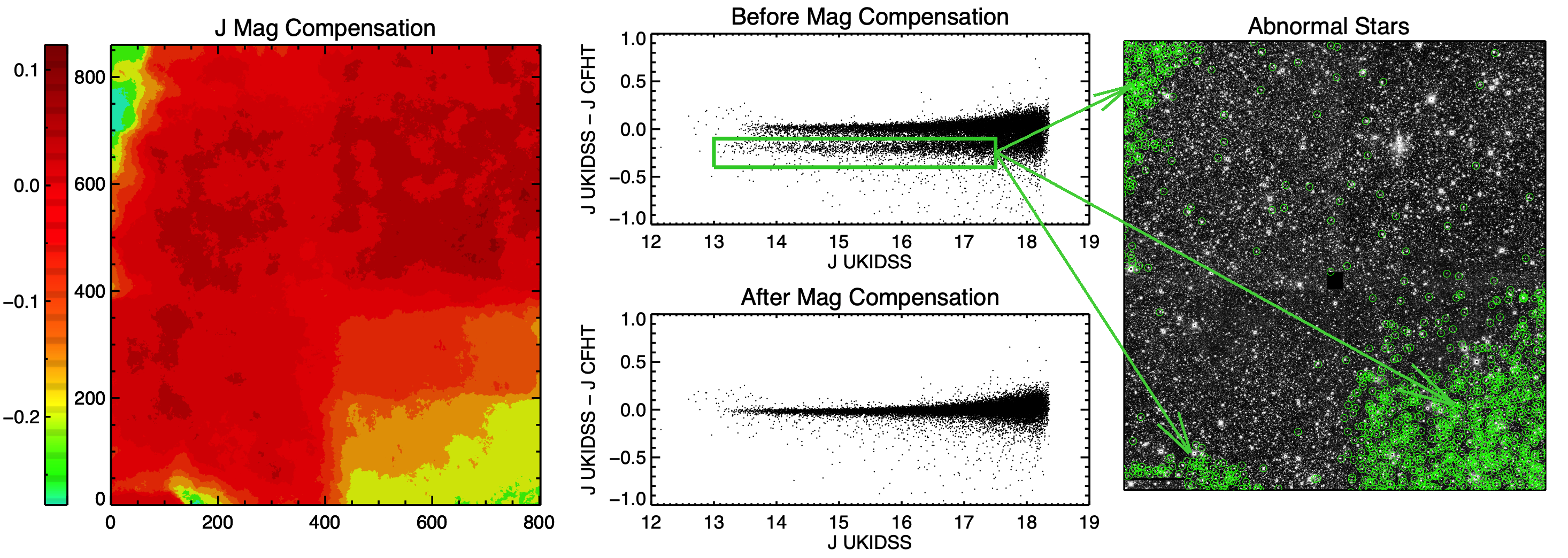}
\caption{
\textit{Left}: J band magnitude ``aperture compensation'' map. Most parts, mainly located in the center, are nearly zero. Those located on edges have bad performance and need correction.  
\textit{Top Middle}: J band magnitude difference between UKIDSS and CFHT catalog for the ``Control'' field.  The lower branch at the bright end is marked in the green box.
\textit{Bottom Middle}: Same stars as in the top right panel after the compensation map is applied.
\textit{Right}: Stars in the green box of the deviated branch are placed on the ``Control'' field map with green circles.
The compensate map of H and K$_{\rm s}$ is almost same with J. For J, the compensate magnitude varies from -0.23 mag to 0.17 mag with a mean value of $\sim$0.038 mag; for H, the compensate magnitude varies from -0.27 mag to 0.058 mag with a mean value of $\sim$-0.057 mag; for K$_{\rm s}$, the compensate magnitude varies from -0.18 mag to 0.094 mag with a mean value of $\sim$0.025 mag.
\label{fig:mag_offset_demo}}
\end{figure*}

We cross-matched CFHT with UKIDSS to process corrections in the CFHT catalog. 
The cross-match goes in each band separately, and there are 289,000, 451,000, and 458,000 stars cross-matched in J, H, and K$_{\rm s}$ bands respectively.
When examining CFHT and UKIDSS magnitude difference diagrams, we found that the points at the brighter end approach two distinct values other than one value and thus form two branches in the diagram.
The second branch is wider and fainter compared to the first one.  
Fig. \ref{fig:mag_offset_demo} top middle panel shows this effect. 

Fig. \ref{fig:mag_offset_demo} right panel shows that relatively faint stars are mostly distributed on the edges or corners, where the astigmatism is worse than those around the center of the image. 
Astigmatism happens because the telescope has slightly different refractive powers on different meridians. 
After the parallel light enters the optic mirrors, it cannot be concentrated at an accurate point but to a slightly deflected point. 
Distorted stars are stretched with images by SWarp during the previous step discussed in Section \ref{chap:follow-up-reduction}, and they can occupy a bigger area than the undistorted stars.
Under the same aperture, these bigger blobs lose some amount of starlight counts.

As fields 1--10 of the ``Target'' repeat in the same observed pattern as the ``Control'' field, their astigmatism patterns are also the same. 
With no extinction from clouds, the ``Control'' field has enough stars that are randomly distributed. 
Therefore the ``Control'' field can be used to draw a compensation map that can be applied to both itself and fields 1--10 of the ``Target'' field. 
For each point of the compensation map, we find 40 nearest neighbor stars on the ``Control'' field that are both detected by UKIDSS and CFHT. 
The mean magnitude difference of them is calculated and taken to be the value of mag compensation at the point. 
Standard deviation of these differences is also calculated, and is placed on a ``compensation deviation'' map. 
Finally, each star in field 0--10 gets a compensated mag added, and mag uncertainties also get error propagation in the same way from the compensation deviation map. 
The J band aperture ``compensation'' map that we made for the astigmatism correction is shown in the left panel of Fig. \ref{fig:mag_offset_demo}.
After this correction, the bright end of magnitude difference between CFHT and UKIDSS converges, as the bottom middle panel of Fig. \ref{fig:mag_offset_demo} shows.

\subsection{Color Term Fitting Correction}\label{chap:colorterm}
Each telescope has its own unique filter transmission curve, and therefore, we need a color term to correlate observations of the same source through CFHT and UKIDSS.
Besides, UKIDSS uses K band, while CFHT uses K$_{\rm s}$ band which is $\sim$25\% narrower.
Given that UKIDSS is a large-scale NIR survey covering a range of areas and depths, it is easier for us to shift the CFHT catalog to the UKIDSS color system.
The 2MASS catalog is not shifted to the UKIDSS color, because only bright stars from 2MASS are combined. 
They are usually saturated in UKIDSS, and lack detections to finish the fittings.

By fitting a linear term with the median value of each segment, transformation equations are as follows. 
First, for stars detected in at least two bands of CFHT, we use equations:
\begin{gather}
J_\mathrm{CFHT} - J_\mathrm{UKIDSS} = a_J + b_J (J_\mathrm{CFHT} - H_\mathrm{CFHT}), \\
H_\mathrm{CFHT} - H_\mathrm{UKIDSS} = a_H + b_H (H_\mathrm{CFHT} - K_\mathrm{sCFHT}),\\
K_\mathrm{sCFHT} - K_\mathrm{UKIDSS} = a_K + b_K (H_\mathrm{CFHT} - K_\mathrm{sCFHT}).
\end{gather}
In the first case we use CFHT magnitudes instead of UKDISS at the right parts, because stars have to be detected in two bands to use these equations.
The qualified stars in UKIDSS are around half of in CFHT (174,000 vs. 376,000 for both J and H detected; 248,000 vs. 510,000 for both H and K). 
The relationships between $J_{\rm CFHT} - H_{\rm CFHT}$ and $J_{\rm UKDISS} - H_{\rm UKIDSS}$, $H_{\rm CFHT}$ - $K_{\rm CFHT}$ and $H_{\rm UKIDSS}$-$K_{\rm UKIDSS}$ are checked to be linear, therefore these equations are compliant with those using UKIDSS at the right part.

Second, for stars detected only in one band, or stars lack key band in the first case, we use $A_{\rm K}$ as the color term for fitting instead:
\begin{gather}
J_\mathrm{CFHT} - J_\mathrm{UKIDSS} = a_J + b_{J}A_{K},\\
H_\mathrm{CFHT} - H_\mathrm{UKIDSS} = a_H + b_{H}A_{K},\\
K_\mathrm{sCFHT} - K_\mathrm{UKIDSS} = a_K + b_{K}A_{K}.
\end{gather}
The second case is for the stars that are significantly different from UKIDSS, lack detections and meanwhile locate in a high extinction place. Their magnitude difference will be reduced through fitting with its $A_{K}$ value. 
$A_{\rm K}$ data are derived from Herschel column density map and converted using $N_{\rm H}/A_{\rm V}=1.8\times10^{21} \rm cm^{-2}/mag$ \citep{predehl1995} and $A_{\rm K}/A_{\rm V}=0.112$ \citep{rieke1985}.
It is roughly a total extinction by all the molecular clouds along the line of sight, and is used as $A_{\rm K}$ TOTAL in the Section \ref{chap:ysos_and_MC}. 
The relationships between $J_{\rm CFHT} - H_{\rm CFHT}$, $H_{\rm CFHT} - K_{\rm CFHT}$ and $A_{K}$ are also checked to be approaching linear, which makes the second case to be compliant with the first.

\begin{figure*}
\centering
\includegraphics[width=0.95\linewidth]{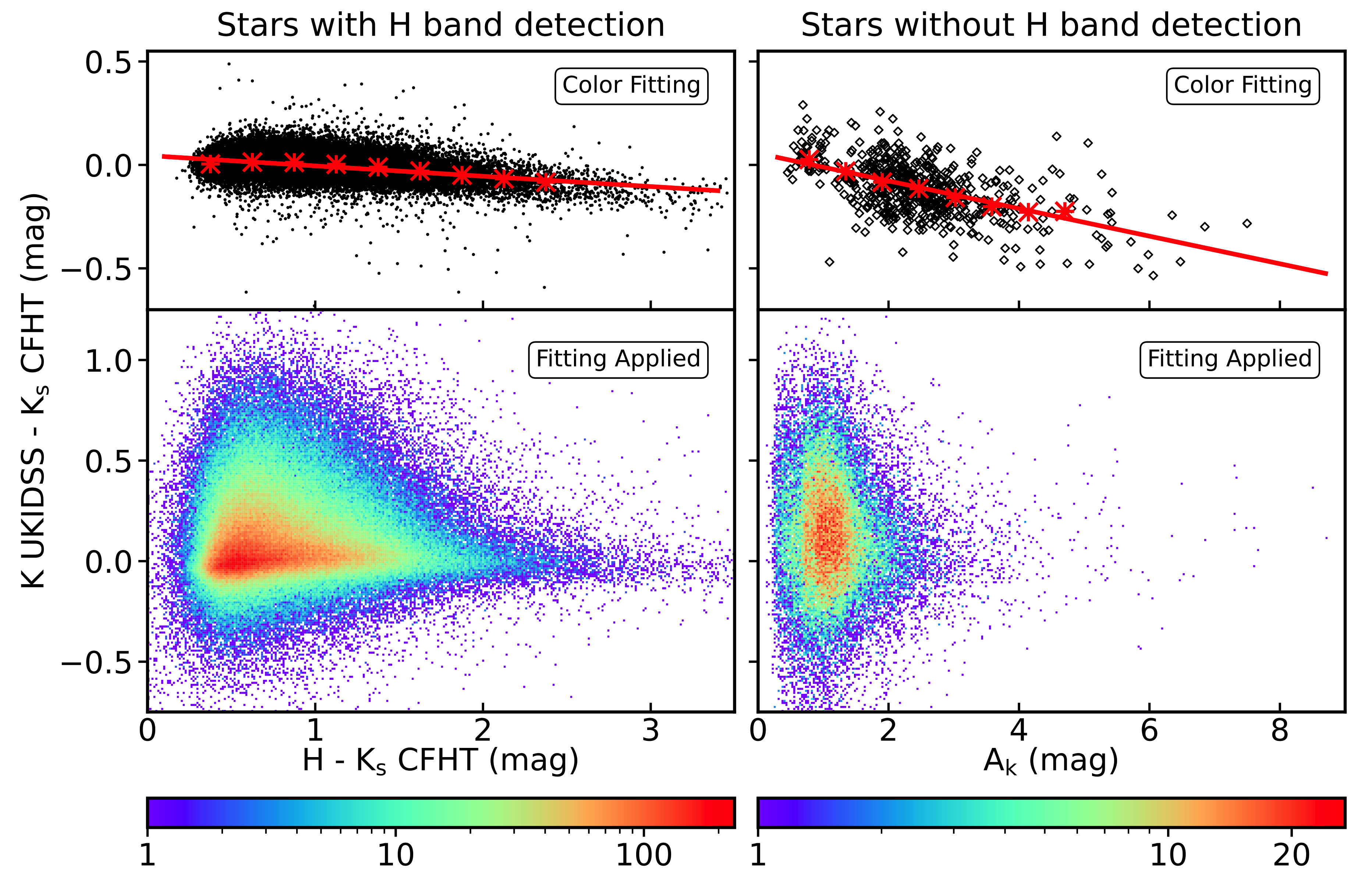}
\caption{
\textit{Left}: Color term fitting of K$_{\rm s}$ band magnitude with H $-$ K$_{\rm s}$ (Top), and after it is applied to the CFHT catalog (Bottom). 
\textit{Right}: Color term fitting with $A_{\rm K}$ (Top), and after it is applied to all qualified stars (Bottom). The median values of each magnitude bin, i.e. the red asterisks, are the actual fitted points. 
Color bar of the bottom panels are counts of 2D histogram divided by 300 $\times$ 200 pixel grids.
Concerning that the color correction is mainly aiming stars with larger $A_{\rm K}$, the red blob in the bottom middle is allowed to be higher than zero here.
These differences with UKIDSS are better corrected in the later zero magnitude correction section.
}\label{fig:color_term_fitting}
\end{figure*}

After the color term is applied, a deeply reddened star has less color difference between CFHT and UKIDSS catalog. 
Fig. \ref{fig:color_term_fitting} shows the magnitude difference of K$_{\rm s}$ band before and after color term fitting. 
Stars that have low magnitude errors ($<$ 0.08), nearest neighbor distance $>$ 10 pixels, and de-reddened K$_{\rm s} <$ 16 mag are used to do the fitting, and the results are applied to all other stars, as bottom two panels of Fig. \ref{fig:color_term_fitting} shows.
For those stars with H-K$_{\rm s}$ $>$ 2.7 mag, K$_{\rm s}$ can get a color correction of around 0.05 mag; for those stars with A$_{\rm K}$ $>$ 4 mag, K$_{\rm s}$ can get a color correction of over 0.2 mag.

\subsection{Zero Magnitude Correction}
\begin{figure}
\centering
\includegraphics[width=0.95\linewidth]{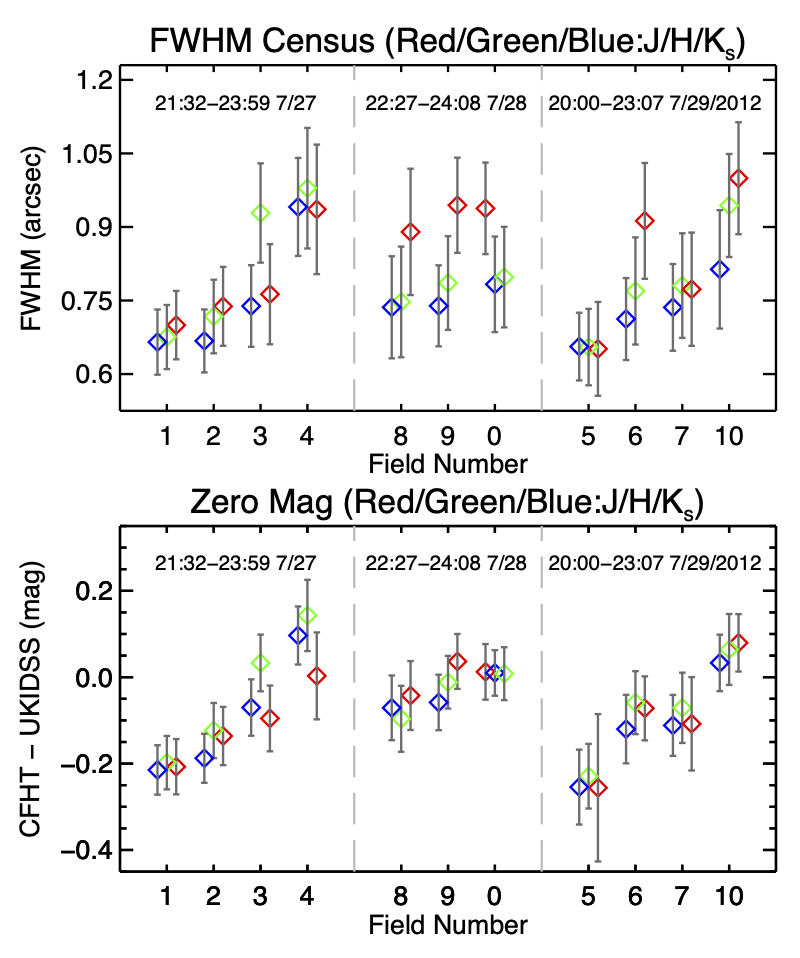}
\caption{
\textit{Top}: FWHM level and variation census of each band in sub-field images. 
\textit{Bottom}: Zero points of magnitude and mag difference variation level of each band in sub-field catalog. We finally add zero magnitude accordingly.
\label{fig:zeromag_fwhm}}
\end{figure}

The observations were carried out through three days, therefore the seeing turned out to be a variable of time.  
Fig. \ref{fig:zeromag_fwhm} shows the zero magnitude correction we adopted in our work.
The top panel shows the FWHM as a function of field number in those three days.  
Fields 1, 2, 3, 4 are observed on 7/27/2012; fields 8, 9, 0 are on 7/28/2012; and Fields 5, 6, 7, 10 are on 7/29/2012. 
As a result, zero magnitude has different levels among the 10 sub-fields inside the ``Target'' field. 
Therefore, we calculate a zero magnitude for each field so that the CFHT stars can be the most close to the UKIDSS.
The zero magnitudes are calculated by fitting.
First, we choose bright stars inside a window of 2 mag that starts from 14.5, 14, and 13.5 mag for J, H, and $K_{\rm s}$ respectively.
Second, their UKIDSS - CFHT magnitudes are fitted with gaussian distribution.
Third, the peak and the error of fitting are taken as the zero magnitude and its uncertainty of this field.
Fig. \ref{fig:zeromag_fwhm}, bottom panel shows zero magnitudes for each band, which will be added back to our data accordingly. 
 
Uncertainty during fitting will also be added to all stars' uncertainties in the way of error propagation.
Magnitude difference between two catalogs should be consistent with their uncertainties. 
2MASS uncertainty is normal, but uncertainty of UKIDSS is underestimated because it only contains a synthetic error term, as \cite{lucas2008} issued. 
Therefore an extra constant term, $C_\mathrm{Unc.UKIDSS}$ is required for expanding the UKIDSS uncertainty.
We correct it by calculating the compromised magnitude difference: $ (Mag_\mathrm{2MASS} - Mag_\mathrm{UKIDSS}) {(Unc_\mathrm{2MASS} ^2 + Unc_\mathrm{UKIDSS} ^2 )}^{-1/2} $. 
For those bright stars, the standard deviation of it should be mostly less than 1. 
In this calculation only those stars that are isolated, round-shaped and not very sharp in shape are used. 
We calculate $C_\mathrm{Unc.UKIDSS}$ to be 0.0265, 0.0347, and 0.0337 in J, H, and K$_{\rm s}$ bands. 
Then this constant term is added back, and the bigger $(Unc_\mathrm{UKIDSS}^2+C_\mathrm{Unc.UKIDSS}^2)^{1/2}$ replaces the smaller $Unc_{UKIDSS}$. 
After this step the two catalogs, 2MASS and UKIDSS are compatible within their uncertainties.
CFHT uncertainty has already increased gradually in previous processes such as astigmatism correction, color term fitting, and zero magnitude adding. 
It is compatible with 2MASS and revised UKIDSS uncertainties.

\subsection{The Super Near-Mid-Infrared Catalog} \label{chap:final_catalog}
After the above steps, a large NIR catalog is generated. 
We cross-match it with Spitzer MIR data and obtained a super near-mid-infrared catalog. 
The point source catalog has a format like Table \ref{tab:YSOs}. 
It contains following information: star id; coordinates; magnitude and mag uncertainty of J/H/K$_{\rm s}$, IRAC Ch1/Ch2/Ch3/Ch4/, MIPS 24$\mu$m; whether this source is detected in 2MASS/UKIDSS/CFHT survey; front extinction (see Section \ref{chap:yso_classification}), and total extinction (see Section \ref{chap:colorterm}).
Selection rules for NIR photometry in multiple catalogs are as follows:
1, if a star has J $<$ 12.5, or H $<$ 12.5, or $K_{\rm s}$ $<$ 12 then its data comes from 2MASS; 
2, if a star has any band of J, H, K$_{\rm s}$ missing then its data comes from CFHT. 
A census of this super near-mid-infrared catalog is shown in Table \ref{tab:catalog}. 

	\begin{table*}
    	\centering
		\caption{Census of the final super near-mid-infrared catalog composition.}
	    \begin{tabular}{|c|c|c|c|c|c|c|c|c|c|}
		\toprule
		  Survey & 2MASS & UKIDSS & CFHT & Ch1 & Ch2 & Ch3 & Ch4 & MIPS & H\&K$_{\rm s}$ \& 1Ch \\
		 \hline 
		  Total & 4,247 & 159,980 & 737,683 & 122,704 & 119,859 & 43,083 & 32,547 & 1,494 & 111,481\\
		 \bottomrule
		 \end{tabular}
 		\label{tab:catalog}
	\end{table*}

\section{Identification and Distribution of YSOs}\label{chap:id_YSOs}

\subsection{Classification of YSOs}\label{chap:yso_classification}
\begin{figure*}
\centering
\includegraphics[width=0.95\linewidth]{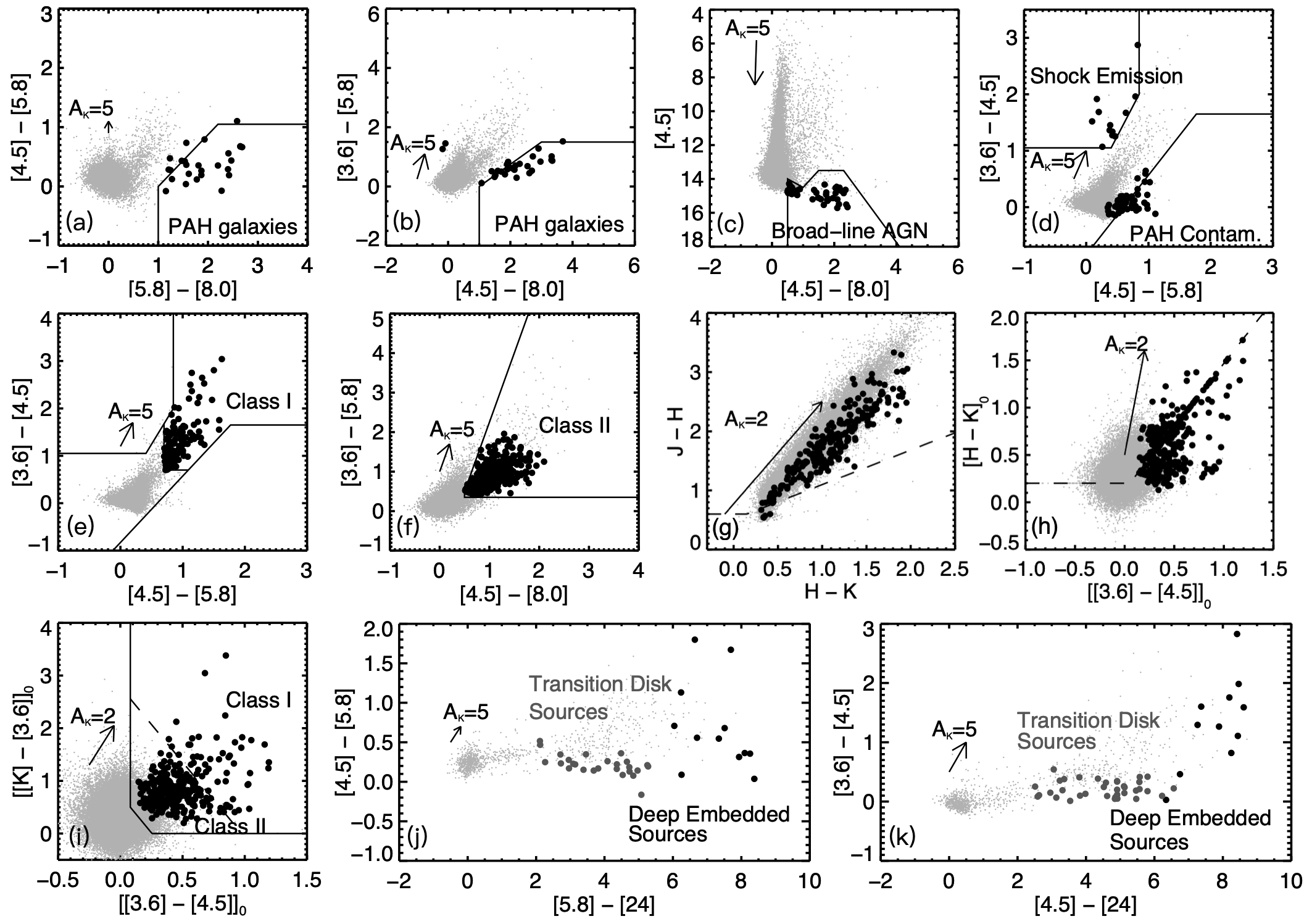}
\caption{ \textit{First Row}: (a--d) Removal of non YSOs sources of star forming galaxies (28), AGNs (47), shock emission (12) and PAH aperture contaminated sources (53).  
\textit{Second Row}: (e, f) Classification of Class I and Class II YSOs with Spitzer Ch1--4 bands.  (g, h) Measuring extinction for dereddening of all the points and dots in (i). (g) shows stars loci before dereddening, and the black dots will be on the two dash lines after the dereddening, like those in (h). Stars not located on the two dash lines in (h) have already been de-reddened in (g). 
\textit{Bottom Row}: (i) Classification of Class I and Class II YSOs with K and Spitzer Ch1--2 bands, and (j, k) classification of deeply embedded sources and transition disk sources with MIPS.  All of the color excess ratios in this figure adopt the extinction law reported by \citet{flaherty2007}.}
\label{fig:Pollutions_YSOs}
\end{figure*}

Based on the super near-mid-infrared catalog, we can identify IR-excess bearing dusty YSOs in the ``Target'' field and generate a comprehensive YSO catalog.
Fig. \ref{fig:Pollutions_YSOs} shows our classification diagrams and processing steps, using the method described in \citet{gutermuth2009} ({\bf G09} hereafter).

First, we remove contamination of non stellar objects, which may have abnormally high fluxes at some specific narrow band channels due to PAH line features, AGN torus emission, or shock heated gas emission, and are confusing for YSO classifications.
Fig. \ref{fig:Pollutions_YSOs} (a--d) show how the distinct color gradient lines are used to separate stars and contamination sources, where the exact expressions of these lines are detailed in G09 appendix A.
Due to the diagrams, We exclude 28 unresolved star forming galaxies, 46 AGNs, 12 shock emission knots, and 49 PAH aperture contaminated sources in the ``Target'' field.
Fig. \ref{fig:Pollu_YSOs_demo}a shows the distribution of these contamination sources.
The shock emission knots tend to appear in places where the molecular clouds have higher column densities, while the other three have roughly random distributions.

Second, we identify and classify Class I/II YSOs according to their infrared excess, mainly by using Spitzer data if all the four IRAC Ch1--4 bands are reliably detected (Fig. \ref{fig:Pollutions_YSOs} e--f).
However, for those that lack IRAC Ch3/4 detections, which turns out to be $\sim$31\% YSOs at last, we recognize them by using the deep K$_{\rm s}$ band data (Fig. \ref{fig:Pollutions_YSOs} i), if their reddening can be measured with at least another NIR band detection.
The dereddening magnitude $A_{\rm K}$ (also called `FRONT' extinction in Section \ref{chap:ysos_and_MC}) is calculated for all those stars with good NIR detections ($\sigma <$ 0.1) demanded in Fig. \ref{fig:Pollutions_YSOs} (g--h). 
The results are listed in the point source catalog in Section \ref{chap:final_catalog}.   
Visually the extinction of a star is proportional to the distance of its locus in Fig. \ref{fig:Pollutions_YSOs} (g--h) to the dash line in the arrow direction of $A_{\rm K}$$=$2.

Third, we search for deeply embedded sources and transition disk sources according to the Spitzer MIPS 24 $\mu$m data (Fig. \ref{fig:Pollutions_YSOs} j--k).
For a deeply embedded source with a lower bolometric temperature or extreme reddening, it is dim in the MIR bands and thus frequently lacks some vital MIR detections or has been misclassified as an AGN contaminant, but is relatively quite bright at MIPS 24 $\mu$m. 
It can either be a Class 0 or a Class I, but cannot be distinguished with current data.
For a transition disk source, it has significant dust clearing within their inner disks, leading to an impression of no disk based on the MIR, but the relatively bright MIPS 24 $\mu$m reveals the presence of the disk beyond the large inferred inner gaps.
It is added to the Class II tallies.
The distributions of these classified YSOs are shown in Fig. \ref{fig:Pollu_YSOs_demo} (b--c).

In total 832 YSOs are classified in the ``Target'' field, including 15 deeply embedded sources, 135 Class I, 647 Class II, and 35 transition disk sources, as shown in Fig. \ref{fig:YSOs_distribution_demo}, and their detailed measurements are listed in Table \ref{tab:YSOs}. 
Besides that, two YSOs are found in the ``Control'' field (one Class II, one transition disk source).
With low column density of molecular clouds and unmatched with YSOs from \cite{dunham2015}, these two are very likely contaminants and are not put in our list.

We have also repeated the same classification procedures by using only 2MASS as the NIR data, and we get 657 YSOs in the ``Target'' field.
With UKIDSS - CFHT data two of the 657 YSOs are not in the final list, and another two get stages re-classified.
Therefore, our UKIDSS - CFHT data finds $\sim$27\% more YSOs.

We briefly cross-matched our list with four former works (with 2$''$ position tolerance) including:  557 YSOs with 2MASS data in all our observed field from \cite{dunham2015}, 264 Class III YSOs with Chandra data in the W40 sub-region from \cite{kuhn2010}, 299 YSOs with Chandra data in the Serpens South sub-region from \cite{winston2018}, and 67 Class 0/I protostars with ALMA data in the Serpens South sub-region from \cite{plunkett2018}. 
Note that Kuhn, Winston, and Plunkett have limited fields of view, and only Dunham has the same field coverage for a full comparison.
We get 379 YSOs matched with \cite{dunham2015} in the ``Target'' field, while get 374 YSOs  matched with the 657 YSOs that are identified using only 2MASS data as described in Section \ref{chap:yso_classification}. 
In the latter case, the other 183 YSOs that are only in Dunham work are randomly distributed, while the 283 only in ours are tracing molecular clouds well.
This is due to the differences between YSOs identification methods of G09 and \cite{harvey2007}.
We match 82 YSOs with \cite{kuhn2010} and 299 YSOs with \cite{winston2018}.
It shows that diskless stars in Serpens South are not as many fractionally as in W40.
We match 33 YSOs with \cite{plunkett2018}, which reveals a fairly younger Serpens South.
In summary, when all these previously identified-YSOs are deducted, 309 are unique in our catalog.

Nevertheless, this YSO catalog may still contain some mis-classified contaminants.
G09 has applied these classification diagrams to the real AGNs that have been identified by \cite{stern2005} in Bootes field.
It suggests a ubiquitous contamination that 7.9 ± 1.0 AGNs will be classified as YSOs per square degree.
Hence it gives us about 8.8 ± 1.1 YSOs to be AGN contaminations in the ``Target'' field (area $\sim$ 1.11 deg$^{2}$).  
On the other hand, some real YSOs can be falsely flagged as PAH emitting sources or AGNs.
\cite{gutermuth2008ngc1333} analyzed a large extragalactic survey field to estimate the residual extragalactic contamination in the YSO census at 6.4 deg$^{-2}$ of Class I sources and 3.8 deg$^{-2}$ of Class II sources.
Thus we might misclassify $\sim$11.3 real YSOs as contamination sources in the ``Target'' field, meanwhile have $\sim$8.8 contamination sources in the YSO list.


\subsection{Spatial distributions of YSOs}
\subsubsection{Distributions of Class I and II YSOs} \label{chap:DC_ysos}

\begin{figure*}
\centering
\includegraphics[width=0.95\linewidth]{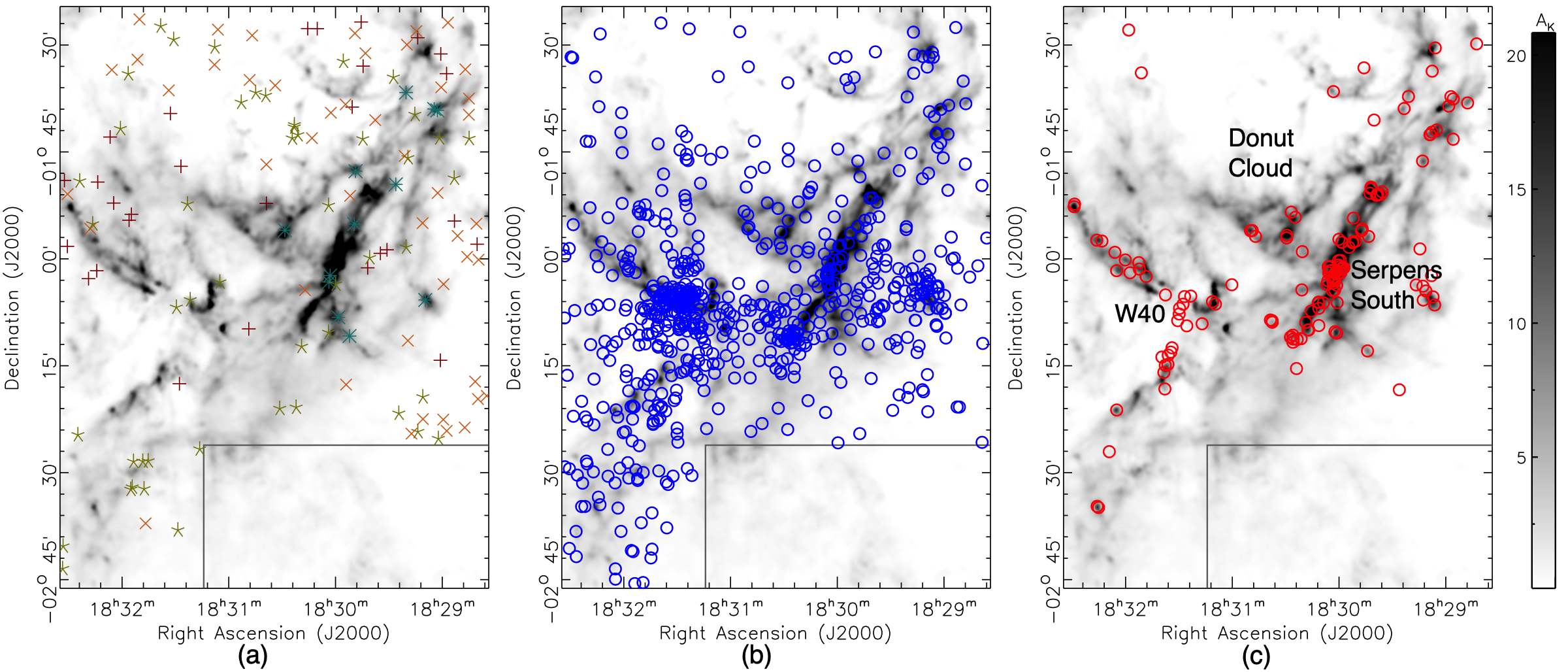}
\caption{
Distributions of contaminations (left), Class II and transition disk sources (middle), Class I and deeply embedded sources (right). 
In the left panel, the positions of PAH galaxies ($+$), AGNs ($\times$), PAH aperture contaminated sources ($\star$), and shock emission knots ($\ast$) are marked. Underlaid images of the three are the same, which is the extinction map of $A_{\rm K}$ derived from Herschel hydrogen column density map.
\label{fig:Pollu_YSOs_demo}}
\end{figure*}

\begin{figure}
\centering
\includegraphics[width=1.05\linewidth]{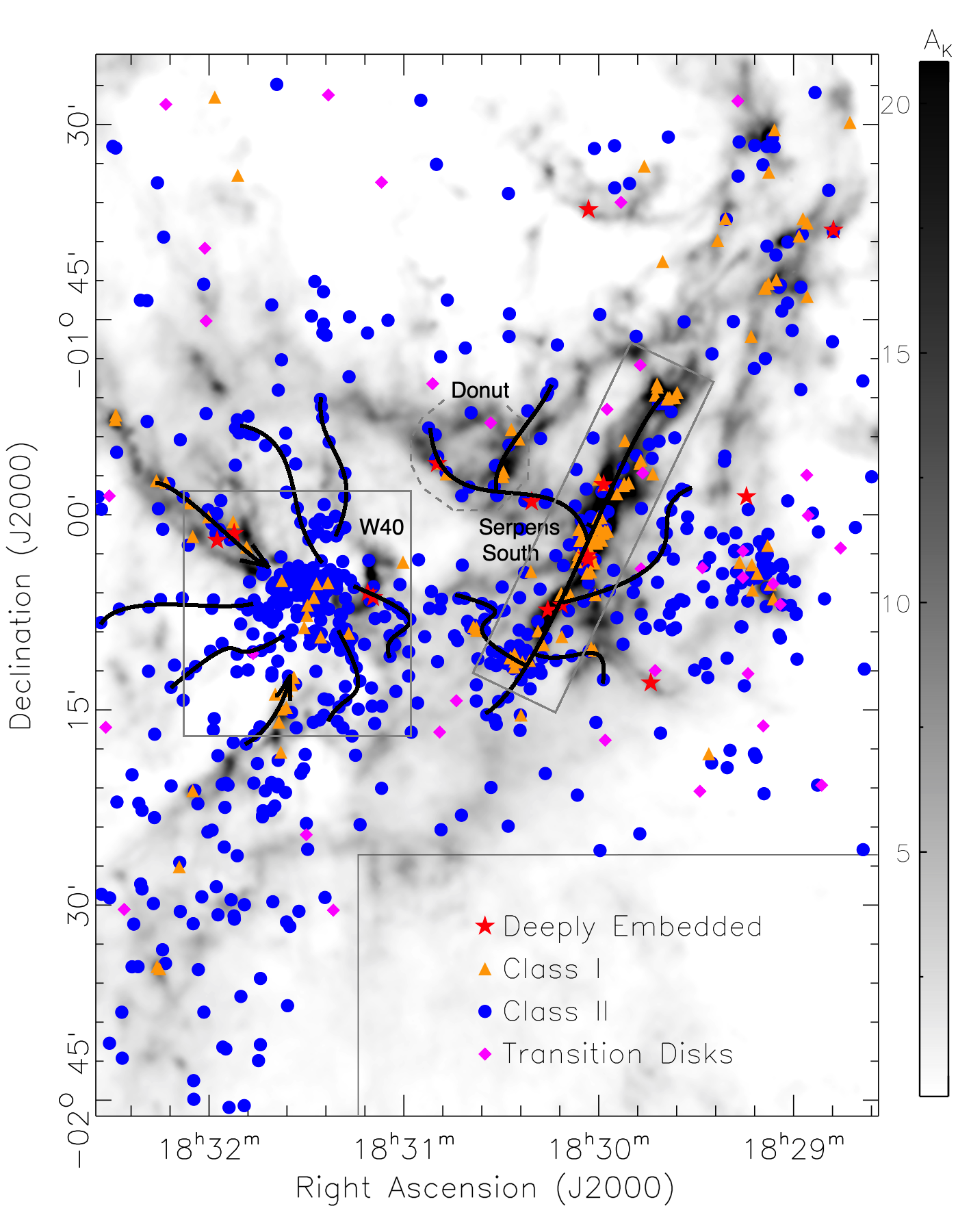}
\caption{
A more detailed YSOs distribution in the ``Target'' field. 
Black curves are the simplifications of nearby MST branches that connect surrounding YSOs to the center cluster. Underlaid image is the same as Fig. \ref{fig:Pollu_YSOs_demo}. The two boxes mark the main parts of W40 and Serpens South clusters. The dash circle marks a ``Donut'' cloud which has a donut-shaped MST subgraph of YSOs.
}
\label{fig:YSOs_distribution_demo}
\end{figure}

\begin{figure}
\centering
\includegraphics[width=\linewidth]{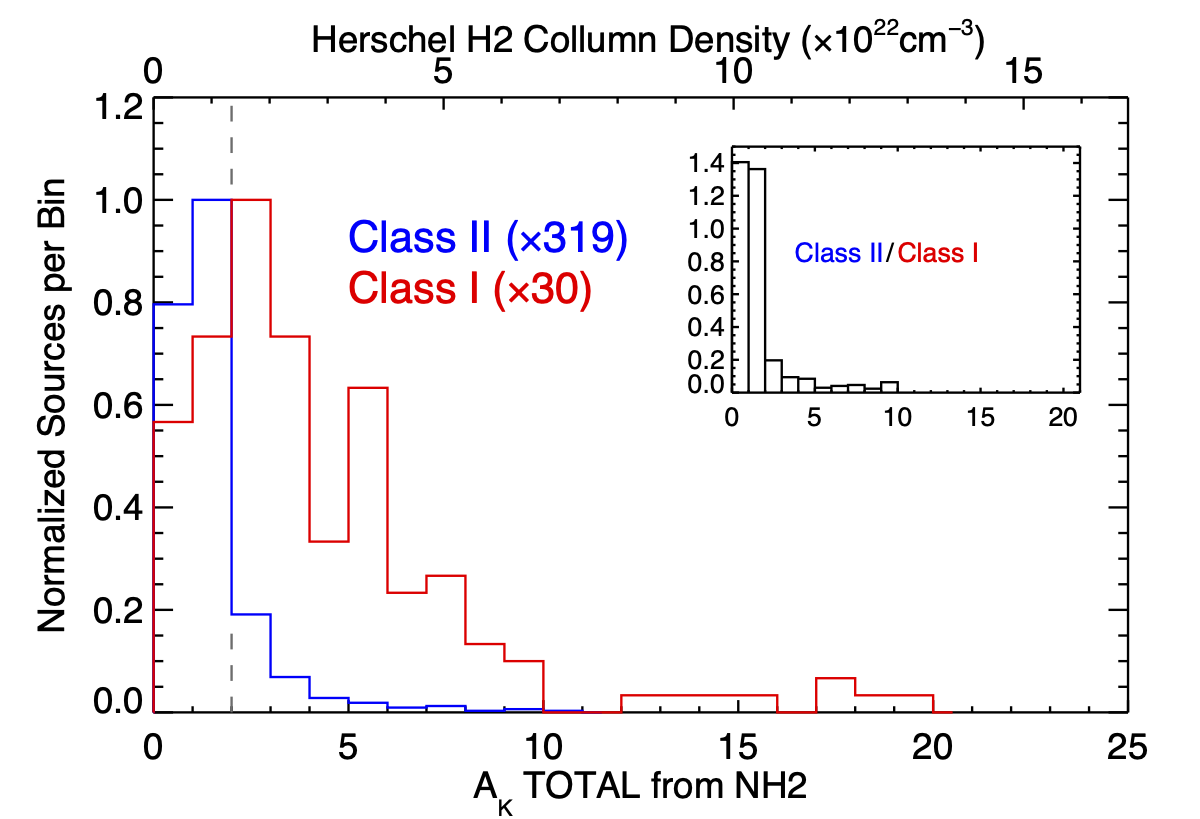}
\caption{
Histogram of total extinction of Class I (including deeply embedded sources) and Class II (including transition disk sources) YSOs. 
Both are normalized by peak bin value, which is 30 for Class I and 319 for Class II. Overlaid dash line marks A$_{\rm K}=2$, where Class I and Class II are almost set apart. The ratio of normalized sources per bin between Class II and Class I are very distinct, and therefore we show them in the subfigure.
\label{fig:Ak_Hist}}
\end{figure}

As shown in Fig. \ref{fig:Pollu_YSOs_demo} (b--c), the Class I YSOs (with deeply embedded sources included) have a stronger correlation with the molecular gas rather than the Class II (with transition disk sources included).
And statistically, Fig. \ref{fig:Ak_Hist} shows the YSO histograms with the column density of molecular clouds in the term of A$_{\rm K}$.
In the entire ``Target'' field, around 74\% Class I YSOs fall into the range of A$_{\rm K} > 2$, but only 16\% Class II YSOs are there.
We can see that most YSOs concentrate into the W40 region and the Serpens South region.

In W40, YSOs are dominated by Class II. 
Based on the existence of O and B MS stars in the central region of W40, \citet{shuping2012} gives an upper limit of the cluster age of about 7 Myrs.
Thus the age of the W40 cluster is in the order of a few Myrs, consistent with typical ages of Class II YSOs.
The two molecular filaments (black curves marked with arrows in Fig. \ref{fig:YSOs_distribution_demo}) that converge to W40 from top left and bottom left contain obviously more Class I YSOs, which are likely outside the W40 HII bubble (green hourglass shaped in Fig. \ref{fig:control_target_demo}).

In the Serpens South region that consists of multiple less massive star forming regions, about a half of YSOs are classified as Class I, with a much higher percentage than that in the W40 region.
Along the central ridge of the molecular filament, massive YSOs are found to have ages mostly less than $\sim$ 0.46--0.72 Myr \citep{plunkett2018}, while a few flat-spectrum YSOs give an upper limit of the cluster age of about 1 Myr \citep{dunham2015}.
And based on the observations of HC7N clumps, the chemical evolution model suggests that the star forming age along the ridge is about less than 0.3 Myr \citep{friesen2013}.
A sub-region north to the Serpens South has its YSOs distributed along the cloud in the shape of a ring, so we name it as ``Serpens South Donut''.
In the Donut region, a HC7N clump is also detected around a deeply embedded source, suggesting a similar young age of about 0.2 Myr.
Consistently, the high percentages of Class I YSOs identified in these two regions are representing a younger cluster age of about less than 1 Myr.

G09 calculated the number ratios of Class I/Class II in 36 star forming cores, and the median value of them is $\sim$0.27. 
The entire W40 - Serpens South region has a similar mean ratio of 0.22.
However, if we take YSOs from two specific box regions in Fig. \ref{fig:YSOs_distribution_demo}, the Class I/Class II ratio is 0.119 for the W40, and 0.849 for the Serpens South.
The Serpens South may rank among the youngest regions \citep{gutermuth2008serpens,li2019}.

\begin{figure}
\centering
\includegraphics[width=0.8\linewidth]{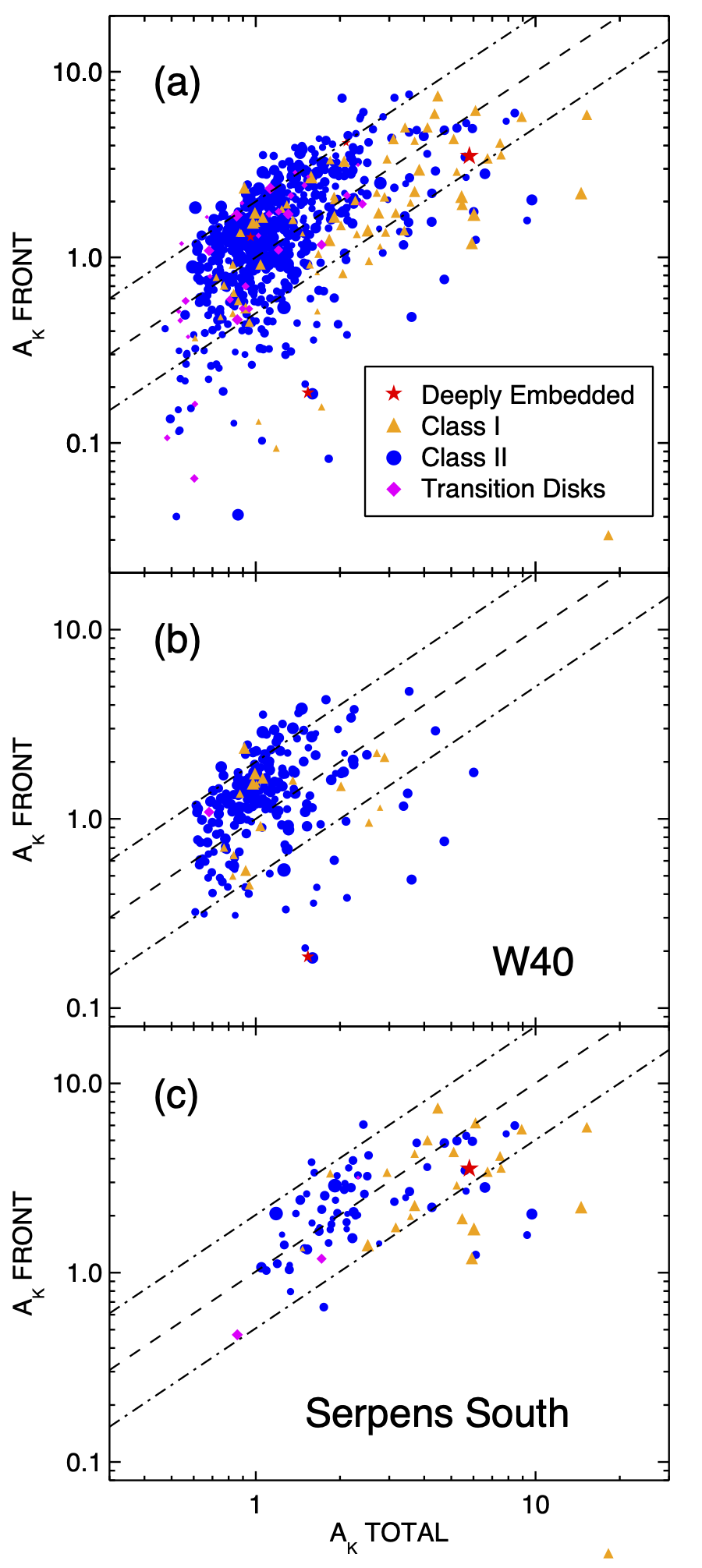}
\caption{
(a) Comparison between the total and the front extinction of all YSOs except for 98 ones that lack NIR detections for A$_{\rm K}$ FRONT and 13 YSOs with A$_{\rm K}$ FRONT$=$0.
(b) and (c) correspond to stars inside the two grey boxes in Fig. \ref{fig:YSOs_distribution_demo}.
The symbol sizes in each species are representing for the relative brightness given by the dereddened magnitude. 
Overlaid dot and dash lines are A$_{\rm K}$ FRONT$=$[2, 1, 0.5] A$_{\rm K}$ TOTAL from top to bottom.
\label{fig:AkvsNh2}}
\end{figure}

\subsubsection{YSOs and Molecular Clouds} \label{chap:ysos_and_MC}

Fig. \ref{fig:AkvsNh2} shows a comparison between the `FRONT' and the `TOTAL' extinction of YSOs, which represents how deeply these YSOs are embedded. 
As stated before, $A_{\rm K}$ FRONT is measured through the NIR/MIR color excess, and $A_{\rm K}$ TOTAL is estimated from the Herschel map. 
For the deeply embedded sources, the magnitude in the MIPS band is used, and the max--min brightness range corresponds to [2.4, 8.6] mag. 
The Class I YSOs use the Ch2 band with a range of [4.7,15.0] mag, and the Class II use the Ch1 band with a range of [3.2, 16.0] mag. 
The transition disk sources use the H band and the range is [6.7,17.3] mag. 
Nevertheless, the `FRONT' extinction of 98 YSOs cannot be measured due to the lack of H band data, and thus they are not on the diagram.

In general, we expect all the points to be located beneath the A$_{\rm K}$ FRONT$=$A$_{\rm K}$ TOTAL equivalent line (the middle black dash line in Fig. \ref{fig:AkvsNh2}), since the gas in front of a star should always be less than the total. 
However, most YSOs in Fig. \ref{fig:AkvsNh2}  are beneath the upper A$_{\rm K}$ line (with a ratio of 2).
Except for the measurement uncertainties of the front and the total extinction, it could also be caused by the different spatial resolutions of telescopes.
Herschel images have a relatively lower resolution of $\sim$20$''$, which can attenuate higher column densities of fibers \citep{hacar2018} to lower averaged values. 
As NIR/MIR images have higher resolution of $\sim$1$''$, the front extinction to YSOs caused by dense fibers is more precisely measured.
Thus the front extinction can be higher than the total value.

Nevertheless, this factor does not prevent us from noticing some interesting phenomena.
(1) The Class II YSOs tend to lay on the region with $A_{\rm K}$ TOTAL $<$ 2 mag, while the Class I appear in the region between 2--20 mag, as also suggested by Fig. \ref{fig:Ak_Hist}.
It indicates the strong correlation between the Class I YSOs and the molecular gas again.
Most Class II YSOs are around the equivalent line, within a factor of two, suggesting that they are lying inside or around the molecular gas.
However, we can also see a fraction of Class II with A$_{\rm K}$ FRONT $<$ 0.6 mag and A$_{\rm K}$ TOTAL $<$ 2 mag are likely having smaller symbol sizes.
These sources can be slightly far away from the molecular filaments, and their locations are hard to form massive YSOs without accreting enough gas.
(2) The YSOs in the W40 region (the square region in Fig. \ref{fig:YSOs_distribution_demo}) are more concentrated around A$_{\rm K} \approx 1$ mag, either for the FRONT or the TOTAL.
These YSOs mainly belong to the compact central cluster that was born in a massive star forming region.
The cluster has been evolved for a few Myrs, and the surrounding molecular gas may be exhausted or expelled by stellar feedback.
It could be the reason why the extinction of these Class II YSOs are relatively lower.
The outliers at higher A$_{\rm K}$ TOTAL roughly lay on the several filamentary branches that still converge toward the W40 cluster.
(3) The YSOs in the Serpens South region (the long rectangular region in Fig. \ref{fig:YSOs_distribution_demo}) are obviously distributed at higher $A_{\rm K}$ TOTAL between 1--20 mag.
Consequently, compared to the Class II, the Class I YSOs are located at the ridge of the filament where the column densities are higher.
Above $A_{\rm K}$ TOTAL $>3$, the $A_{\rm K}$ FRONT of YSOs do not increase linearly, but instead show a flat trend.
It suggests that some YSOs are not presented here, especially those deeply embedded in thick filaments with $A_{\rm K}$ TOTAL $>3$.
But these YSOs are not necessarily undetected.
There are 98 YSOs not in this diagram since lacking the H band detection, and we find 88 of them have $A_{\rm K}$ TOTAL $>2$, in which about 60 are in the Serpens South region.
As a result, most YSOs in this region are likely detected, though some deeply embedded sources do not have the necessary H band for the front extinction calculation.
In addition, while most YSOs are surrounding the $A_{\rm K}$ FRONT$=$A$_{\rm K}$ TOTAL equivalent line, a number of outliers are located below the lower A$_{\rm K}$ line (with a ratio of 0.5) in Fig. \ref{fig:AkvsNh2}. Those outliers also have lower front extinction, although they have higher $A_{\rm K}$ TOTAL values (around 10 mag).
Most of these outliers belong to a small central cluster, including the one at the lower right corner.
(4) The YSOs in the rest regions have a similar situation that the Class I are lying on the thicker filaments.
Most Class II YSOs are around the equivalent line within a factor of two, but a small portion of them and even a few Class I are likely having smaller front extinction, which are usually less bright.
These fainter outliers may be given birth in small star forming regions, where the natal gas is exhausted quickly.

\subsubsection{Minimum Spanning Tree of YSOs}

To characterize the YSO distributions more quantitatively and explore potential connections among different regions, we construct the minimum spanning tree (MST) for these YSOs (e.g., G09).
The MST is a concept in graph theory, but frequently used to separate close stellar clusters \citep[e.g.][]{battinelli1991, hetem1993, cartwright2004, gutermuth2009, kirk2014}. 
The MST here is generated by connecting the nearest neighbor of each point first to form an ``island'', and then connecting every two nearby ``islands'' with the shortest path. 
New ``islands'' gradually grow bigger, until all points get connected to form a tree structure that has the total branch length to be the minimum.

Fig. \ref{fig:YSOs_distribution_demo} shows some prominent MST branches of the main body of the W40 and the Serpens South regions, while the detailed analysis of the MST is presented in the following paper (Sun et al. 2022, submitted to MNRAS).
The Donut cloud is young and its branches are well connected to the Serpens South rather than the W40 region.
Some clusters located in the west or northwest to the Serpens South region are relatively independent, whose MST branches are not marked here.

For W40, eight branches are spreading to all directions from the central cluster, in the shape of a spider. 
Six of the branches, illuminated by Class II YSOs, have no obvious corresponding filamentary molecular clouds.
But the other two match the filaments well, and they contain more Class I YSOs.
Therefore, the W40 central cluster is likely sitting on a converge point of many filaments in the past, as suggested by many works that massive star forming regions need to accrete more gas from multiple sub-filaments \citep[e.g.,][]{schneider2012,peretto2013,rayner2017}.

The Serpens South region also has several filaments connected, in the shape of a gecko, and its main body has been on a thick and over-dense filament.
The Class I YSOs (including the deeply embedded sources) roughly lay on intersection points of the MST branches and the ridge of the main filament.
It also suggests the gas supplying from sub-filaments.
Though the main body is depicted by a long and straight MST branch, it consists of several smaller star forming regions.
Based on the HC7N detections, the star forming age along the ridge of the molecular filament is less than about 0.3 Myr \citep{friesen2013}.
But at the north part of this filament, where the cloud is the most over-dense, the age is even younger.

\section{Summary}\label{chap:sum}

We conduct a super near-and-mid-infrared catalog of point sources in the W40 - Serpens South region, based on the deep NIR data of CFHT observations in combination with 2MASS, UKIDSS, and Spitzer catalogs.
To identify stellar sources thoroughly, the data reduction processes including image stacking and photometry corrections are required to be performed precisely.
For the CFHT data, we adopt the `Astromatic' suite and optimize the reduction processes, instead of using the SIMPLE pipeline, to obtain precise images for photometry.
Multiple corrections are applied to these photometry results, such as that for optics astigmatism and telescope spikes, color term fitting, and zero magnitude correction.
To maximize the data accuracy, we use 2MASS and UKIDSS as the standard catalog to resolve saturated bright stars, subdivide the zero magnitude, and perform the astigmatism correction. 
Catalogs are cross calibrated before being integrated together. 
In addition, we cross-match the results with Spitzer data, and include the publicly available Herschel hydrogen column density map to demonstrate the total dust extinction.
At last, all the information is combined to form this super near-mid-infrared catalog of point sources for this region.

Based on this super catalog, we identify 832 young stellar objects (YSOs) after filtering contamination sources out, and classify 15, 135, 647, and 35 of them to be the deeply embedded sources, the Class I, Class II YSOs, and transition disk sources. 
We also compare the front extinction of these YSOs with the total column densities at their locations to check their depths in the molecular clouds, and generate the MST of the YSOs to check their spatial correlation with the molecular clouds.
In general, the YSOs are well correlated with the filamentary structures of molecular clouds in this region, especially the Class I YSOs (including the deeply embedded sources).
The W40 central region is dominated by the evolved Class II YSOs, and the surrounding materials were likely expelled by previous stellar feedback.
Around the W40 central cluster, there are eight prominent MST branches, in which six branches exclusively populated with Class II YSOs show little correlation with molecular filaments, but the other two include detectable gas and corresponding Class I YSOs.
The YSOs in the Serpens South region are mainly laying on the thick filamentary clouds, and the Class I YSOs roughly emerge at intersection points where the sub-branches join with the main body filament.
The front extinction of them is also higher compared to that in the W40 region, and the H band emission of some deeply embedded ones can even not penetrate the thick molecular gas.
In summary, the massive star formation in the W40 - Serpens South region is commonly supported by gas supply from multiple sub-filaments.
Our results imply a convoluted YSO distribution in this important star forming region.

\section*{Acknowledgements}
We sincerely appreciate the anonymous referee for the very careful reading and helpful comments, which greatly improved the manuscript.
We gratefully thank all the memberships in the PMO star forming group for their valuable discussions.
M. Zhang is supported by the National Natural Science Foundation of China (NSFC grant No.12073079).
H. Wang and Y. Ma acknowledge the support by NSFC grants No.11973091 and No.11973090.
S.N. Zhang acknowledges the support from NSFC grant No.11573070.
R.A. Gutermuth gratefully acknowledges funding support for this work from NASA ADAP award NNX17AF24G.
J. Sun acknowledges financial support from the China Scholarship Council (CSC). 
J. Sun sincerely thanks Chi-hung Yan for his valuable explanation on CFHT `I'wii pipeline.
Observations of this work are obtained with WIRCam, a joint project of CFHT, Taiwan, Korea, Canada, France and the Canada–France–Hawaii Telescope (CFHT) which is operated by the National Research Council (NRC) of Canada, the Institute National des Sciences de l’Univers of the Centre National de la Recherche Scientifique of France and the University of Hawaii. 
This work is based in part on observations made with the Spitzer Space Telescope, which was operated by the Jet Propulsion Laboratory, California Institute of Technology under a contract with NASA.
This research has made use of data from the Herschel Gould Belt survey (HGBS) project, which is a Herschel Key Programme jointly carried out by SPIRE Specialist Astronomy Group 3 (SAG 3), scientists of several institutes in the PACS Consortium, and scientists of the Herschel Science Center (HSC). 
This research has made use of the data products from the UKIDSS Galactic Plane Survey (GPS), which is one of the five NIR Public Legacy Surveys that are being undertaken by the UKIDSS consortium, and is available online through the WFCAM Science Archive (WSA). 
This research has made use of the VizieR catalog access tool, CDS, Strasbourg, France. 
Through the VizieR we make use of data products from the Two Micron All Sky Survey, which is a joint project of the University of Massachusetts and the Infrared Processing and Analysis Center/California Institute of Technology, funded by the National Aeronautics and Space Administration and the National Science Foundation.
This research has made use of astronomical pipeline softwares from the ``AstrOmatic'' which is located at Institut d’Astrophysique de Paris (IAP), France, and benefits from support by French PNC and PNG.
This research has made use of TOPCAT, which is an interactive graphical viewer and editor for tabular data.

\bibliographystyle{mnras}
\bibliography{MNscript_w40-sS.bbl} 

\begin{landscape}
 \begin{table}
		\caption{The near-mid-infrared catalog of identified YSOs in ``Target'' field. Two default values, -100 and 10 are used as placeholders for non-detections in magnitude and uncertainty. The full table is available online.}\label{tab:YSOs}
	    \scriptsize
	    \setlength{\tabcolsep}{3pt}
	    \begin{tabular}{ccccccccccccccccccccccc} 
\toprule
ID & RA (J2000) & DEC (J2000) & J & unc.J & H & unc.H & K & unc.K & NIR Catalog & 3.6  & unc.3.6 & 4.5 & unc.4.5 & 5.8 & unc.5.8 & 8.0 & unc.8.0 & 24 & unc.24 & $A_{K} FRONT$ & $A_{K} TOTAL
$ & Class\\
		\hline
   & hh\ mm\ ss.sss & dd\ mm\ ss.sss & mag & mag & mag & mag & mag & mag & mag & mag  & mag & mag & mag & mag & mag & mag & mag & mag & mag & mag & mag & (I*,I,II,II*)\\
		\hline
   \multicolumn{23}{c}{$Deeply\ Embedded\ Sources$}\\
		\hline
       1&18:30:3.115s&-1:36:32.770s&  -100.000&    10.000&    17.840&     0.029&    16.194&     0.019&CFHT&    13.653&     0.019&      12.388  &     0.011&    11.842&     0.024&    11.521&     0.060&     4.500&     0.005&     0.000&     2.223&I*\\
       2&18:28:47.771s&-1:38:7.869s&  -100.000&    10.000&  -100.000&    10.000&    17.021&    10.000&CFHT&    13.509&     0.011&      11.524  &     0.010&    11.160&     0.009&    10.346&     0.017&     3.062&     0.002&  -100.000&     2.452&I*\\
       3&18:31:57.414s&-2:1:57.435s&    17.005&     0.030&    16.227&     0.036&    16.168&     0.047&UKIDSS&    16.134&     0.254&      15.026  &     0.137&    13.227&     0.357&    11.461&     0.346&     6.587&     0.191&     0.187&     1.536&I*\\
       4&18:31:52.180s&-2:1:26.148s&    22.972&     0.973&  -100.000&    10.000&    17.782&    10.000&CFHT&    13.632&     0.017&      11.876  &     0.007&    11.196&     0.019&    10.728&     0.080&     3.685&     0.024&  -100.000&     3.358&I*\\
       5&18:30:3.259s&-2:3:26.575s&  -100.000&    10.000&  -100.000&    10.000&  -100.000&    10.000&NONE&  -100.000&    10.000&    -100.000  &    10.000&  -100.000&    10.000&    10.414&     0.040&     4.400&     0.040&  -100.000&    12.796&I*\\
		
	  ... & & & & & & & & & & & & & & & & & & & \\
      14&18:30:15.645s&-2:7:19.774s&  -100.000&    10.000&    18.288&     0.032&    15.600&     0.022&CFHT&    13.207&     0.016&      12.388  &     0.012&    12.074&     0.057&    11.647&     0.066&     4.141&     0.006&     3.518&     5.800&I*\\
      15&18:29:43.972s&-2:12:55.408s&  -100.000&    10.000&  -100.000&    10.000&    17.493&    10.000&CFHT&    14.203&     0.025&      12.612  &     0.015&    12.256&     0.042&    12.279&     0.079&     4.003&     0.002&  -100.000&     2.396&I*\\
		\hline
		\multicolumn{23}{c}{$Class\ I\ YSOs $}\\
		\hline	 
      16&18:31:51.010s&-1:33:54.766s&    19.490&     0.127&    17.431&     0.046&    15.625&     0.039&UKIDSS&    13.503&     0.013&      12.641  &     0.008&    11.777&     0.017&    10.852&     0.014&     7.286&     0.032&     0.781&     0.723&I\\
      17&18:31:58.086s&-1:27:53.997s&    19.945&     0.065&    18.232&     0.036&    16.694&     0.022&CFHT&    14.562&     0.030&      13.764  &     0.025&    12.964&     0.041&    11.901&     0.034&     7.852&     0.031&     0.477&     0.748&I\\
	  
	  ... & & & & & & & & & & & & & & & & & & & \\
     149&18:32:15.980s&-2:34:43.280s&    18.352&     0.018&    15.503&     0.009&    13.003&     0.019&CFHT&    11.795&     0.004&      10.181  &     0.004&     9.334&     0.003&     8.780&     0.010&     3.310&     0.001&     1.394&     3.395&I\\
     150&18:32:9.134s&-2:27:3.158s&  -100.000&    10.000&    17.652&     0.021&    16.355&     0.016&CFHT&    14.781&     0.099&      13.945  &     0.065&  -100.000&    10.000&  -100.000&    10.000&  -100.000&    10.000&     0.130&     1.023&I\\
	  
		\hline
		\multicolumn{23}{c}{$Class\ II\ YSOs $}\\
		\hline	 
     151&18:31:24.861s&-1:46:1.704s&    10.697&     0.023&     8.870&     0.053&     7.756&     0.026&2MASS&     6.809&     0.002&       6.592  &     0.002&     6.313&     0.001&     6.018&     0.001&     5.121&     0.005&     1.047&     0.739&II\\
     152&18:31:24.681s&-1:45:20.551s&    12.394&     0.023&    10.488&     0.022&     9.652&     0.019&2MASS&     8.841&     0.002&       8.527  &     0.002&     7.930&     0.002&     6.893&     0.002&     4.713&     0.006&     1.375&     0.714&II\\
	  
	  ... & & & & & & & & & & & & & & & & & & & \\
     796&18:32:20.591s&-2:28:45.903s&    18.827&     0.020&    17.110&     0.014&    16.170&     0.015&CFHT&    15.074&     0.061&      14.424  &     0.038&  -100.000&    10.000&  -100.000&    10.000&  -100.000&    10.000&     1.033&     1.013&II\\
     797&18:32:21.005s&-2:28:22.727s&    18.549&     0.020&    16.638&     0.012&    15.700&     0.013&CFHT&    14.761&     0.038&      14.199  &     0.032&  -100.000&    10.000&  -100.000&    10.000&  -100.000&    10.000&     1.342&     0.920&II\\
	  	  
		\hline
		\multicolumn{23}{c}{$Transition\ Disk\ Sources$}\\
		\hline	
     798&18:32:0.791s&-1:45:6.379s&    17.958&     0.038&    15.224&     0.035&    13.746&     0.034&UKIDSS&    12.846&     0.007&      12.697  &     0.008&    12.464&     0.025&    12.406&     0.060&     9.739&     0.191&     2.150&     1.336&II*\\
     799&18:32:1.153s&-1:39:31.191s&  -100.000&    10.000&    17.681&     0.017&    16.401&     0.016&CFHT&    14.573&     0.034&      14.334  &     0.037&  -100.000&    10.000&    12.199&     0.040&     9.206&     0.073&     1.963&     0.893&II*\\

		  ... & & & & & & & & & & & & & & & & & & & \\
     831&18:31:21.688s&-2:30:24.095s&    11.869&     0.028&     9.429&     0.024&     7.951&     0.027&2MASS&     6.400&     0.002&       6.293  &     0.002&     5.777&     0.001&     5.527&     0.001&     3.664&     0.003&     1.684&     0.854&II*\\
     832&18:32:26.095s&-2:30:20.597s&    17.363&     0.015&    15.061&     0.007&    13.852&     0.015&CFHT&    13.096&     0.008&      12.896  &     0.011&    12.712&     0.034&    12.792&     0.103&     7.987&     0.109&     1.713&     1.116&II*\\
\bottomrule
  \end{tabular}
 \end{table}
\end{landscape}

\label{lastpage}
\end{document}